\documentclass[aps,prb,preprint,showpacs,showkeys,floatfix]{revtex4}
\usepackage{graphicx,color}
\newcommand{\rev}[1]{\textcolor{black}{#1}}
\graphicspath{{figs/}}
\begin{document}

\title{Buckled Graphene for Efficient Energy Harvest, Storage, and Conversion}

\author{Jin-Wu Jiang}
    \altaffiliation{jwjiang5918@hotmail.com}
    \affiliation{Shanghai Institute of Applied Mathematics and Mechanics, Shanghai Key Laboratory of Mechanics in Energy Engineering, Shanghai University, Shanghai 200072, People's Republic of China}

\date{\today}
\begin{abstract}

Buckling is one of the most common phenomena in atomic-thick layered structures like graphene. While the buckling phenomenon usually causes disaster for most nano-devices, we illustrate one positive application of the buckled graphene for energy harvest, storage, and conversion. More specifically, we perform molecular dynamical simulations to show that the buckled graphene can be used to collect the wasted mechanical energy and store the energy in the form of internal knotting potential. Through strain engineering, the knotting potential can be converted into useful kinetic (thermal) energy that is highly concentrated at the free edges of the buckled graphene. The present study demonstrates potential applications of the buckled graphene for converting the dispersed wasted mechanical energy into the concentrated useful kinetic (thermal) energy.

\end{abstract}
\keywords{Buckled Graphene, Energy Harvest, Energy Storage, Energy Conversion}
\pacs{72.20.mq, 62.25.-g, 61.48.Gh}
% 72.20.mq, Buckling
% 62.25.-g, Mechanical properties of nanoscale systems 
% 61.48.Gh, Structure of graphene
\maketitle
%\tableofcontents
\pagebreak

%\section{Introduction}

Graphene is a quasi two-dimensional (2D) honeycomb lattice structure of high in-plane stiffness\cite{LeeC2008sci} but very small bending modulus.\cite{OuyangZC1997,TuZC2002,ArroyoM2004,LuQ2009} As a result of the quasi 2D nature, buckling becomes the most common phenomenon in graphene. For the buckling instability, Euler buckling theory\cite{TimoshenkoS1987} states that the critical compression strain, above which graphene will be buckled, is inversely proportional to the in-plane stiffness $C_{11}$ and is proportional to the bending modulus $D$; i.e., $\epsilon_c \propto D/C_{11}$. According to the Euler buckling theory, the critical strain for graphene is very small; i.e., the buckling phenomenon can easily take place in graphene. Consequently, the buckling process can be induced by very weak external disturbance like the thermal expansion effect.\cite{BaoW2009nn}

The buckling of graphene has attracted intensive research interests in past few years.\cite{LuQ2009ijam,PatrickWJ2010jctn,SakhaeePA2009cms,PradhanSC2009cms,PradhanSC2009plsa,FrankO2010acsnn,FarajpourA2011pe,TozziniV2011jpcc,RouhiS2012pe,GiannopoulosGI2012cms,Neek-AmalM2012apl,ShenH2013apl} In most of these existing works, buckling brings negative effects on the mechanical, thermal, or electronic properties of graphene. However, in the present work, we will show that buckled graphene can collect the wasted mechanical energy and convert the wasted energy into useful concentrated kinetic energy.

It has been found that, the conversion of energy on the nanoscale level plays an important role in supporting the engineering of nano-devices. Chang performed molecular dynamics simulations to examine the Domino-like energy transformation between the van der Waals potential and the kinetic (thermal) energy in single-walled carbon nanotubes.\cite{ChangT2008prl} Many works have proposed to use graphene-based materials as flexible supercapacitors for energy storage and conversion.\cite{StollerMD2008nl,WangY2009jpcc,LiuC2010nl,WangZ2011adm,BrownsonDAC2011jps} 

In this paper, we demonstrate a mechanical route for the application of graphene for energy collection, storage, and conversion. More specifically, the buckled graphene can be used to collect wasted mechanical energy, which is stored in the form of knotting potential. The energy stored in the buckled graphene can be converted into kinetic energy concentrated at the free edges of graphene. We also investigate possible methods to increase the efficiency for the energy conversion from the dispersed mechanical energy into the concentrated kinetic energy.

\begin{figure}[tb]
  \begin{center}
    \scalebox{1.0}[1.0]{\includegraphics[width=8cm]{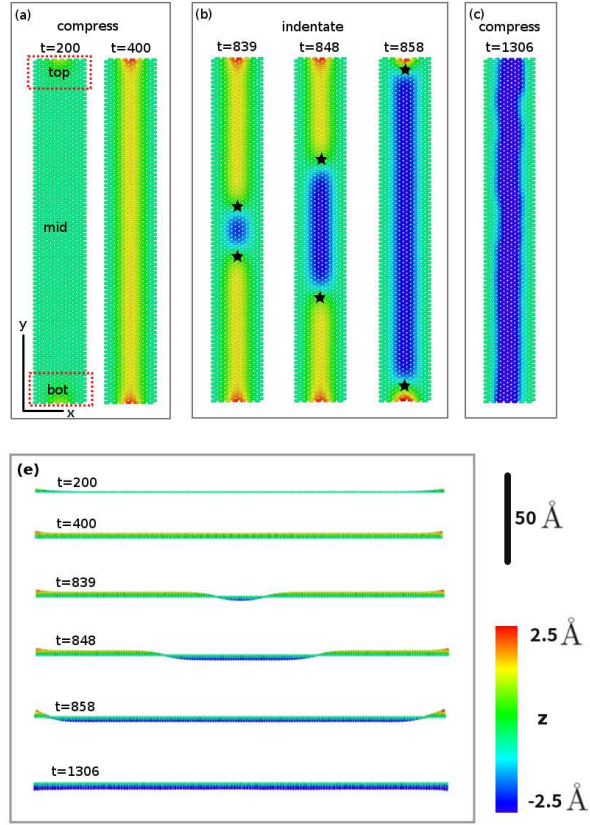}}
  \end{center}
  \caption{(Color online) Energy harvest, storage, and conversion using buckled graphene of dimension $30\times 200$~{\AA} at 1~K. (a) Graphene is buckled under external mechanical compression along the x-direction. (b) The buckled graphene is indented at the middle position, creating two knots (depicted by black stars), which move to the $\pm$y ends at t=858~ps. These knots are stable at the free edges of the buckled graphene. This step is to mimic the collection of the wasted mechanical energy (mimic by moving indenter), which is transformed into the potential energy of the knots. (c) The knots are loosened by further compression along the x-direction, which leads to the conversion of the knoting potential energy into the kinetic energy concentrated at the $\pm$y edges. (e) Side views for the five configurations. Color is with respective to the z-coordinate of each atom.}
  \label{fig_md_process}
\end{figure}

\begin{figure}[tb]
  \begin{center}
    \scalebox{1.0}[1.0]{\includegraphics[width=8cm]{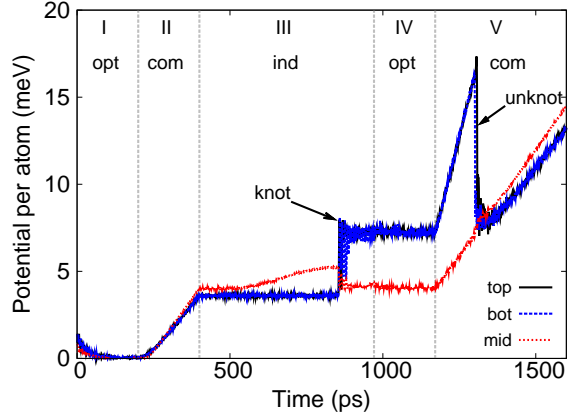}}
  \end{center}
  \caption{(Color online) Potential per atom for the top (black solid line), bottom (blue dashed line), and middle (red dotted line) regions during the energy collection, storage, and conversion in graphene of dimension $30\times 200$~{\AA}. There are five typical steps for the simulation. (I) Graphene is thermalized within the NPT ensemble for 200~ps. (II) Graphene is buckled by compression along the x-direction. (III) The buckled graphene is indented and two knots are created. (IV) The knotting graphene is thermalized within the NPT ensemble for another 200~ps. (V) The knots are loosened by further compression along the x-direction.} 
  \label{fig_potential}
\end{figure}

\begin{figure}[tb]
  \begin{center}
    \scalebox{1.0}[1.0]{\includegraphics[width=8cm]{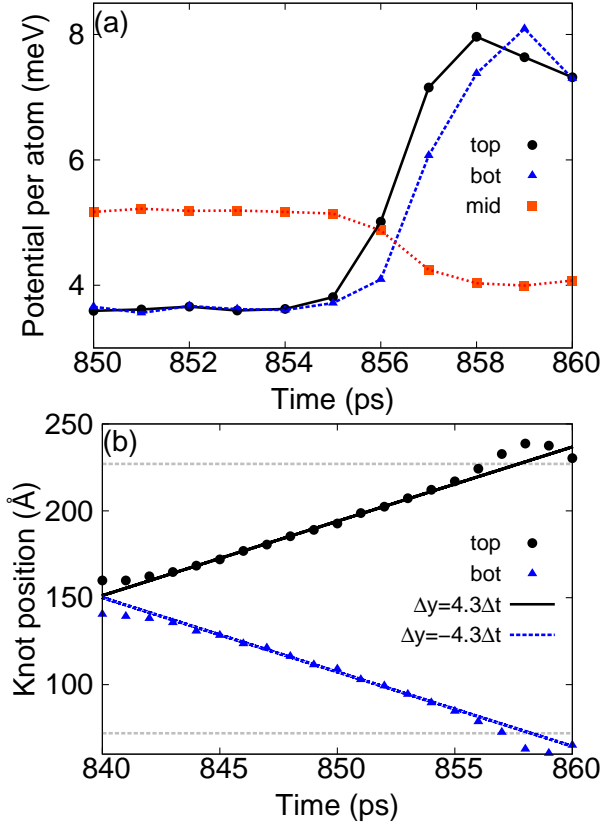}}
  \end{center}
  \caption{(Color online) The motion of the knots created in step III in Fig.~\ref{fig_potential}. (a) A close-up of the potential per atom during the motion of the knots. Both knots are in the middle region before t=856~ps, after which one knot moves to the top region and the other knot moves to the bottom region. (b) The time dependence for the position of the knots. These knots travel at a speed of 430~{ms$^{-1}$}. Two horizontal dashed lines depict the boundary for the two edge regions in graphene along the y-direction.} 
  \label{fig_knot_position}
\end{figure}

\begin{figure}[tb]
  \begin{center}
    \scalebox{1.0}[1.0]{\includegraphics[width=8cm]{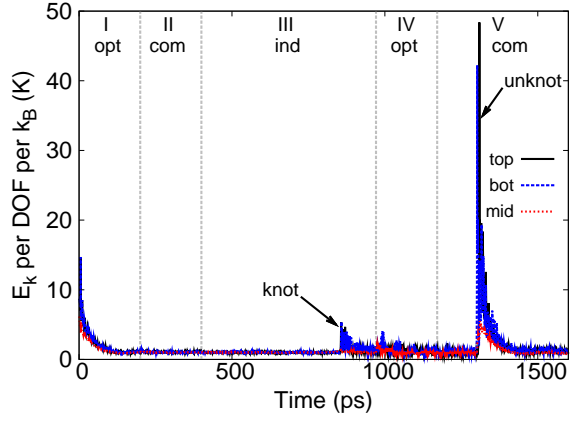}}
  \end{center}
  \caption{(Color online) The evolution of the total kinetic energies for the top, bottom, and middle regions in the graphene of dimension $30\times 200$~{\AA}. The y-axis shows the total kinetic energy divided by the total degrees of freedom for each region.} 
  \label{fig_kinetic}
\end{figure}

\begin{figure}[tb]
  \begin{center}
    \scalebox{1.0}[1.0]{\includegraphics[width=8cm]{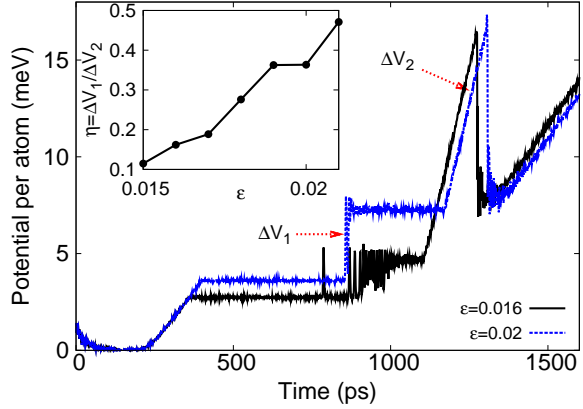}}
  \end{center}
  \caption{(Color online) The potential energy per atom for the top region in graphene, which is buckled by compression with strain magnitude $\epsilon=0.016$ and 0.02. The dimension of the graphene is $30\times 200$~{\AA}. The potential variation during the knot formation is $\Delta V_1$, while the potential variation during the loosening is $\Delta V_2$. The efficiency ($\eta=\Delta V_1/\Delta V_2$) of the energy collection is displayed in the inset.} 
  \label{fig_potential_strain}
\end{figure}

\begin{figure}[tb]
  \begin{center}
    \scalebox{1.0}[1.0]{\includegraphics[width=8cm]{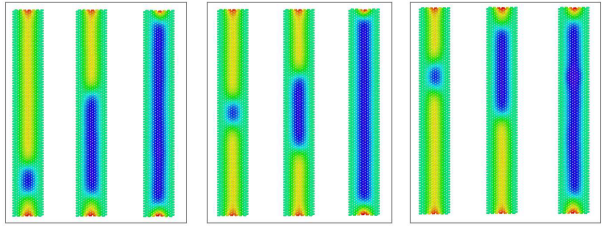}}
  \end{center}
  \caption{(Color online) The energy conversion process for buckled graphene which is indented at position $y_{\rm ind}=$ $\frac{1}{6}$, $\frac{1}{2}$, and $\frac{2}{3}$ (with respective to the width in the y-direction). The dimension of the graphene is $30\times 200$~{\AA}. Color indicates the atomic z-coordinate.} 
  \label{fig_md_pos}
\end{figure}

\begin{figure}[tb]
  \begin{center}
    \scalebox{1.0}[1.0]{\includegraphics[width=8cm]{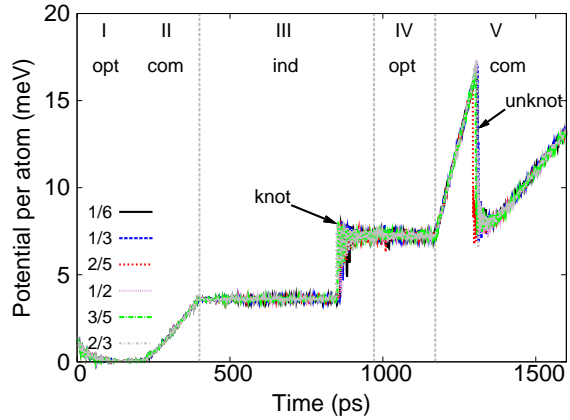}}
  \end{center}
  \caption{(Color online) The potential energy per atom for the top region in buckled graphene, which is indented at $y_{\rm ind}=$ $\frac{1}{6}$, $\frac{1}{3}$, $\frac{2}{5}$, $\frac{1}{2}$, $\frac{3}{5}$, and $\frac{2}{3}$ (with respective to the width in the y-direction). The dimension of the graphene is $30\times 200$~{\AA}.} 
  \label{fig_potential_pos}
\end{figure}

\begin{figure}[tb]
  \begin{center}
    \scalebox{1.0}[1.0]{\includegraphics[width=8cm]{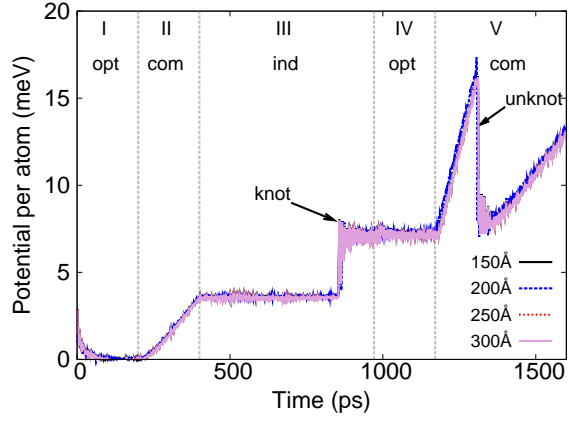}}
  \end{center}
  \caption{(Color online) The potential per atom for the top region in graphene of width 150, 200, 250, and 300~{\AA} in the y-direction. The length is 30~{\AA} in the x-direction.} 
  \label{fig_potential_width}
\end{figure}

\begin{figure}[tb]
  \begin{center}
    \scalebox{1.0}[1.0]{\includegraphics[width=8cm]{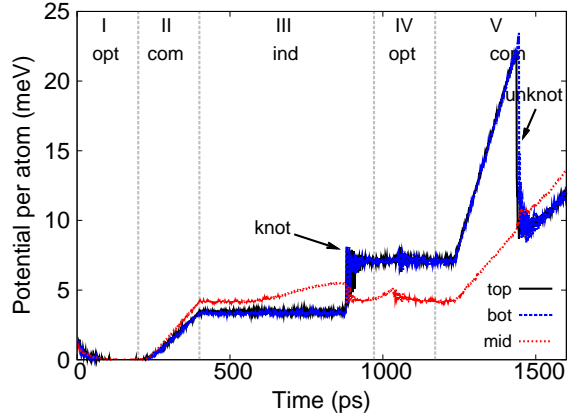}}
  \end{center}
  \caption{(Color online) Potential per atom for the top (black solid line), bottom (blue dashed line), and middle (red dotted line) regions during the energy collection, storage, and conversion in zigzag graphene of dimension $30\times 200$~{\AA}.} 
  \label{fig_potential_zigzag}
\end{figure}

\begin{figure}[tb]
  \begin{center}
    \scalebox{1.0}[1.0]{\includegraphics[width=8cm]{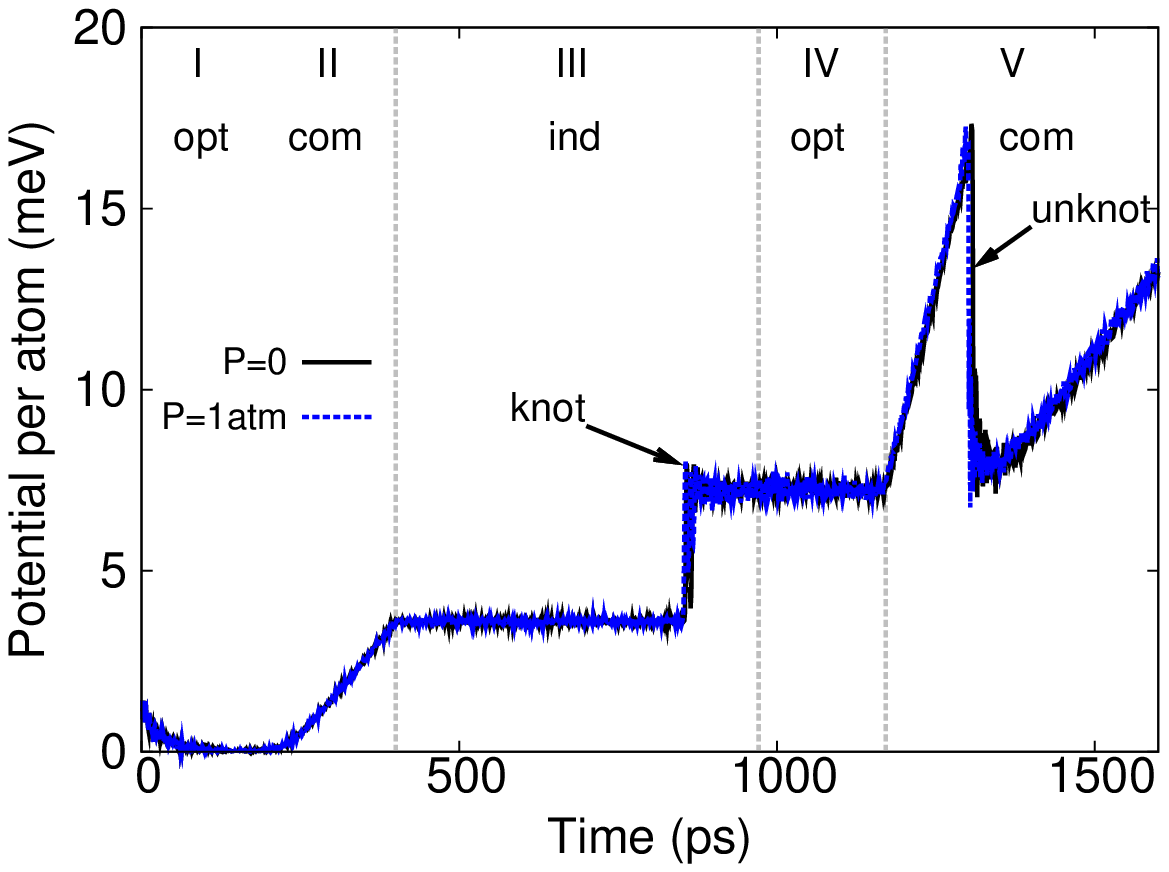}}
  \end{center}
  \caption{(Color online) The potential per atom for the top region in graphene of dimension $30\times 200$~{\AA} at pressure 1~atm and 0~atm.} 
  \label{fig_potential_pressure}
\end{figure}

%\section{Simulation details}

The left structure in Fig.~\ref{fig_md_process}~(a) shows the thermalized configuration for graphene of dimension $30\times 200$~{\AA}. Both ends in the x-direction are fixed, while free boundary conditions are applied in y and z-directions. The whole system is divided into the top, middle (mid), and bottom (bot) regions. Free edges are at the top and bottom regions. The interactions between carbon atoms in graphene are described by the second generation Brenner potential.\cite{brennerJPCM2002} The standard Newton equations of motion are integrated in time using the velocity Verlet algorithm with a time step of 1~{fs}. The Nos\'e-Hoover\cite{Nose,Hoover} thermostat is used for maintaining the constant temperature at 1~K and the constant pressure of 0~Pa. Molecular dynamics simulations are performed using the publicly available simulation code LAMMPS.\cite{PlimptonSJ} The OVITO package is used for visualization.\cite{ovito}

Fig.~\ref{fig_md_process} displays the whole energy collection, storage, and conversion process in five simulation steps. First, the system is thermalized to the targeted pressure and temperature within the NPT (i.e. the particles number N, the pressure P and the temperature T of the system are constant) ensemble for 200~ps. Second, graphene is buckled by compression along the x-direction for 200~ps at a strain rate of $10^{-4}$~{ps$^{-1}$}, which results in a final compressive strain of 2\% in the system. The structure is allowed to be fully relaxed in lateral directions during mechanical loading. The buckled structure is shown by the right configuration in Fig.~\ref{fig_md_process}~(a).

In the third step, the buckled graphene is indented by a spherical indenter tip. The indenter is moving toward the buckled graphene at a constant speed of 0.01~{\AA ps$^{-1}$}. Fig.~\ref{fig_md_process}~(b) shows that, during the indentation process, two knots (indicated by stars) are created in the middle region and these knots will move to the top and bottom regions in the buckled graphene. This step is to mimic the collection of the wasted mechanical energy. More specifically, the buckled graphene is hit by the indenter (with wasted mechanical energy) at the central position, which leads to the formation of two knots, i.e., the wasted mechanical energy can be collected and stored as the potential for these two knots (indicated by stars). We note that the third structure shown in Fig.~\ref{fig_md_process}~(b) is very stable, which indicates that buckled graphene can store energy in the form of potential energies for knots near the edge.

In the fourth step, the buckled graphene with two knots are thermalized within the NPT ensemble for another 200~ps. In the fifth step, the system is compressed again in the x-direction, and the knots will be loosened eventually. As a result, the knotting potential energy is released as kinetic energy, most of which concentrates at the free edges of graphene. The fifth step is to mimic the usage of the knoting potential through mechanical engineering.

%\section{Results and discussions}
We now illustrate the whole process by examining the potential energy and kinetic energy during these five simulation steps. Fig.~\ref{fig_potential} shows the potential per atom for graphene during the whole simulation process, which indicates that the wasted mechanical energy is transformed into the knotting potential in the middle region at t=600~ps. These knots move to the top and bottom regions at $t\approx856$~ps as shown in Fig.~\ref{fig_knot_position}~(a), at which the potential of the top and bottom regions increases suddenly while the potential of the middle region decreases. The potential energies of these two knots are recorded in Fig.~\ref{fig_knot_position}~(b), giving a speed of 430~{ms$^{-1}$} for the motion of the knot. Both knots are loosened at $t\approx 1300$~ps, when the potential energies of the top and bottom regions decrease suddenly as shown in Fig.~\ref{fig_potential}~(a).

The evolution of the total kinetic energy is shown in Fig.~\ref{fig_kinetic}. We focus on the time around 1300~ps, when the kinetic energies of the top and bottom regions show a sudden increase, which indicates the occurrence of the loosening phenomenon. The change of the kinetic energy in the middle region is much smaller than the kinetic energy variations in the top or bottom region. It means that the knotting potential is converted into the kinetic energy for atoms in the edge (top and bottom) region. From this simulation step, we learn that the wasted mechanical energy is eventually converted into the concentrated kinetic (thermal) energy, which is useful for engineering nano-devices.

We further examine possible effects from using different simulation parameters. In the above second simulation step, graphene is buckled by compressive strain $\epsilon$ along the x-direction. We find that the magnitude of the strain has considerable effect on the efficiency of the energy conversion. Fig.~\ref{fig_potential_strain} shows that the potential variation during the knot formation is $\Delta V_1$, which can be regarded as the external mechanical energy that can be collected by the buckled graphene. The potential variation during the loosening is $\Delta V_2$, which is the work to be done to utilize the stored energy. The efficiency of the energy conversion can be defined as $\eta=\Delta V_1/\Delta V_2$. With the increase of $\epsilon$, the potential variation during knotting ($\Delta V_1$) increases, while $\Delta V_2$ decreases. Hence, the efficiency increases with increasing $\epsilon$ as shown in the inset of Fig.~\ref{fig_potential_strain}. The efficiency for the energy conversion can be as high as 47\% for $\epsilon=0.021$. This efficiency is obviously larger than the energy conversion efficiency of solar cells, which is typically lower than 30\%.\cite{GreenMA2014ppra}

Fig.~\ref{fig_md_pos} shows that the position of the indenter during the indentation process is not important. The buckled graphene is indented at different positions, but the same knoting structure is formed at the end. Fig.~\ref{fig_potential_pos} shows that the evolution of the potential per atom is almost the same for the buckled graphene that are indented at different positions. The position insensitivity is a nice property for energy collection in sense that the buckled graphene can collect the wasted mechanical energy that is dispersed in the space. Fig.~\ref{fig_potential_width} shows that the width of graphene has no effect on the whole energy collection and conversion processes either. However, the energy harvest/storage/conversion processes are dependent on the creation of knots, that can be stable at the edges of the graphene ribbons. The knots are created at the middle of the graphene ribbon, and the knots will travel to and stay at the free edges. Thus a proper width to length ratio is required for the knots to be stable at the edges of the graphene ribbon.

The above graphene ribbons are of armchair orientation along the x-direction.  We show in Fig.~\ref{fig_potential_zigzag} that the orientation of the graphene ribbon is not important for the performance of the energy device. We found similar energy collection, storage, and conversion processes in the graphene of zigzag orientation in the x-direction.

Considering the large Young's modulus (about 1.0~{TPa}) for graphene, the internal stress (pressure) is usually several orders larger than the ambient pressure of 101.3~{kPa}. More specifically, for a typical strain $\epsilon=0.01$ used in the present work, the internal stress is $\sigma=E\epsilon=1.0TPa\times 0.01=10~{GPa}$, which is five orders larger than the ambient pressure. Hence, the ambient pressure is neglectable. This speculation is verified in Fig.~\ref{fig_potential_pressure}, which shows that the potential curve for pressure $P=1$~{atm} is almost the same as the potential curve for $P=0$.

We note that the working temperature for the energy collection and conversion based on the buckled graphene is related to the intrinsic energy scale in this energy device. More specifically, we have introduced the parameter $\eta=\Delta V_1/\Delta V_2$ as the efficiency for the energy device, with $\Delta V_1$ as the potential variation during the knot formation and $\Delta V_2$ as the potential variation during unknotting. That is $\Delta V_1$ is the energy that can be collected by the buckled graphene, while $\Delta V_2$ is the external work to be done to explore the energy stored in the device. In other words, $\Delta V_2$ serves as a potential barrier for the knot. Hence, the working temperature should be lower than $T_C=\Delta V_2/k_B$. Otherwise, the knot can be loosened by the thermal vibration, i.e., the knot becomes thermally unstable, if the thermal vibration energy is larger than the potential barrier $\Delta V_2$. From Fig.~\ref{fig_potential_pressure}, we have $\Delta V_2\approx 10$~{meV}, so the critical temperature is about $T_C\approx 116$~{K}. We have checked that the knot indeed becomes thermally unstable at room temperature, so the working temperature for the energy device should be lower than 116~K. \rev{From Fig.~\ref{fig_potential_strain}, the potential variation $\Delta V_2$ will increase with increasing strain $\epsilon$, so the critical temperature $T_C$ can be increased by increasing strain $\epsilon$. However, as shown in the above, the efficiency for the energy device will be decreased with the increase of $\Delta V_2$, so there is a trade-off between higher critical temperature and higher efficiency in the real-world application of the buckled graphene for energy harvest/storage/conversion.}

\rev{Finally, the energy harvest and storage will be valuable for some physical processes, especially on the nanoscale level. As an example, we propose one possible application of the buckled graphene for energy harvest/storage/conversion in nanomechanical resonators, which can work at temperatures from room temperature down to 10~K.\cite{ChenC2009nn,ZandeAMVD} The resonant oscillation energy will be decayed into wasted mechanical energy after a long time, which can be harvested by the buckled graphene discussed in the present work. The harvested energy will be stored in the buckled graphene as knotting potential. We have found in our work that the stored energy can be converted into the thermal vibration that is highly localized at the free edges, which will be useful for the actuation of some localized resonant oscillations of the nanomechanical resonators.\cite{SanchezDG}}

Overall, we have demonstrated in the above that the buckled graphene can be used to collect wasted mechanical energy. The dispersed mechanical energy can be collected through hitting the buckled graphene at various positions, which leads to the same final knotting configuration. The buckled graphene is long in the y-direction, and this large surface area is helpful for energy harvest. Furthermore, the collected energy can be converted into the kinetic energy concentrated at the two free edges, which may be useful for mechanical engineering of nano-devices.

%\section{conclusion}
In conclusion, we have investigated the application of buckled graphene to collect dispersed mechanical energy. The mechanical energy is stored within the buckled graphene in the form of knotting potential, which can be utilized as the kinetic energy localized at the free edges of graphene. One advanced feature for the energy conversion process using buckled graphene is that such system is able to collect mechanical energy that is highly dispersed in the space, and convert the energy into highly localized kinetic (thermal) energy.

\textbf{Acknowledgements} The author thanks Tien-Chong Chang and Xing-Ming Guo at SHU for helpful discussions. The work is supported by the Recruitment Program of Global Youth Experts of China, the National Natural Science Foundation of China (NSFC) under Grant No. 11504225 and the start-up funding from Shanghai University.

%\bibliographystyle{aipnum4-1}
%\bibliographystyle{nature}
%\bibliography{/home/JiangJinWu/Documents/papers/mypapers/latex/biball}

\begin{thebibliography}{34}%
\makeatletter
\providecommand \@ifxundefined [1]{%
 \@ifx{#1\undefined}
}%
\providecommand \@ifnum [1]{%
 \ifnum #1\expandafter \@firstoftwo
 \else \expandafter \@secondoftwo
 \fi
}%
\providecommand \@ifx [1]{%
 \ifx #1\expandafter \@firstoftwo
 \else \expandafter \@secondoftwo
 \fi
}%
\providecommand \natexlab [1]{#1}%
\providecommand \enquote  [1]{``#1''}%
\providecommand \bibnamefont  [1]{#1}%
\providecommand \bibfnamefont [1]{#1}%
\providecommand \citenamefont [1]{#1}%
\providecommand \href@noop [0]{\@secondoftwo}%
\providecommand \href [0]{\begingroup \@sanitize@url \@href}%
\providecommand \@href[1]{\@@startlink{#1}\@@href}%
\providecommand \@@href[1]{\endgroup#1\@@endlink}%
\providecommand \@sanitize@url [0]{\catcode `\\12\catcode `\$12\catcode
  `\&12\catcode `\#12\catcode `\^12\catcode `\_12\catcode `\%12\relax}%
\providecommand \@@startlink[1]{}%
\providecommand \@@endlink[0]{}%
\providecommand \url  [0]{\begingroup\@sanitize@url \@url }%
\providecommand \@url [1]{\endgroup\@href {#1}{\urlprefix }}%
\providecommand \urlprefix  [0]{URL }%
\providecommand \Eprint [0]{\href }%
\providecommand \doibase [0]{http://dx.doi.org/}%
\providecommand \selectlanguage [0]{\@gobble}%
\providecommand \bibinfo  [0]{\@secondoftwo}%
\providecommand \bibfield  [0]{\@secondoftwo}%
\providecommand \translation [1]{[#1]}%
\providecommand \BibitemOpen [0]{}%
\providecommand \bibitemStop [0]{}%
\providecommand \bibitemNoStop [0]{.\EOS\space}%
\providecommand \EOS [0]{\spacefactor3000\relax}%
\providecommand \BibitemShut  [1]{\csname bibitem#1\endcsname}%
\let\auto@bib@innerbib\@empty
%</preamble>
\bibitem [{\citenamefont {Lee}\ \emph {et~al.}(2008)\citenamefont {Lee},
  \citenamefont {Wei}, \citenamefont {Kysar},\ and\ \citenamefont
  {Hone}}]{LeeC2008sci}%
  \BibitemOpen
  \bibfield  {author} {\bibinfo {author} {\bibfnamefont {C.}~\bibnamefont
  {Lee}}, \bibinfo {author} {\bibfnamefont {X.}~\bibnamefont {Wei}}, \bibinfo
  {author} {\bibfnamefont {J.~W.}\ \bibnamefont {Kysar}}, \ and\ \bibinfo
  {author} {\bibfnamefont {J.}~\bibnamefont {Hone}},\ }\href@noop {} {\bibfield
   {journal} {\bibinfo  {journal} {Science}\ }\textbf {\bibinfo {volume}
  {321}},\ \bibinfo {pages} {385} (\bibinfo {year} {2008})}\BibitemShut
  {NoStop}%
\bibitem [{\citenamefont {Ou-Yang.}, \citenamefont {bin Su},\ and\
  \citenamefont {Wang}(1997)}]{OuyangZC1997}%
  \BibitemOpen
  \bibfield  {author} {\bibinfo {author} {\bibfnamefont {Z.-C.}\ \bibnamefont
  {Ou-Yang.}}, \bibinfo {author} {\bibfnamefont {Z.}~\bibnamefont {bin Su}}, \
  and\ \bibinfo {author} {\bibfnamefont {C.-L.}\ \bibnamefont {Wang}},\
  }\href@noop {} {\bibfield  {journal} {\bibinfo  {journal} {Physical Review
  Letters}\ }\textbf {\bibinfo {volume} {78}},\ \bibinfo {pages} {4055}
  (\bibinfo {year} {1997})}\BibitemShut {NoStop}%
\bibitem [{\citenamefont {Tu}\ and\ \citenamefont {Ou-Yang}(2002)}]{TuZC2002}%
  \BibitemOpen
  \bibfield  {author} {\bibinfo {author} {\bibfnamefont {Z.-C.}\ \bibnamefont
  {Tu}}\ and\ \bibinfo {author} {\bibfnamefont {Z.-C.}\ \bibnamefont
  {Ou-Yang}},\ }\href@noop {} {\bibfield  {journal} {\bibinfo  {journal}
  {Physical Review B}\ }\textbf {\bibinfo {volume} {65}},\ \bibinfo {pages}
  {233407} (\bibinfo {year} {2002})}\BibitemShut {NoStop}%
\bibitem [{\citenamefont {Arroyo}\ and\ \citenamefont
  {Belytschko}(2004)}]{ArroyoM2004}%
  \BibitemOpen
  \bibfield  {author} {\bibinfo {author} {\bibfnamefont {M.}~\bibnamefont
  {Arroyo}}\ and\ \bibinfo {author} {\bibfnamefont {T.}~\bibnamefont
  {Belytschko}},\ }\href@noop {} {\bibfield  {journal} {\bibinfo  {journal}
  {Physical Review B}\ }\textbf {\bibinfo {volume} {69}},\ \bibinfo {pages}
  {115415} (\bibinfo {year} {2004})}\BibitemShut {NoStop}%
\bibitem [{\citenamefont {Lu}, \citenamefont {Arroyo},\ and\ \citenamefont
  {Huang}(2009)}]{LuQ2009}%
  \BibitemOpen
  \bibfield  {author} {\bibinfo {author} {\bibfnamefont {Q.}~\bibnamefont
  {Lu}}, \bibinfo {author} {\bibfnamefont {M.}~\bibnamefont {Arroyo}}, \ and\
  \bibinfo {author} {\bibfnamefont {R.}~\bibnamefont {Huang}},\ }\href@noop {}
  {\bibfield  {journal} {\bibinfo  {journal} {Journal of Physics D: Applied
  Physics}\ }\textbf {\bibinfo {volume} {42}},\ \bibinfo {pages} {102002}
  (\bibinfo {year} {2009})}\BibitemShut {NoStop}%
\bibitem [{\citenamefont {Timoshenko}\ and\ \citenamefont
  {Woinowsky-Krieger}(1987)}]{TimoshenkoS1987}%
  \BibitemOpen
  \bibfield  {author} {\bibinfo {author} {\bibfnamefont {S.}~\bibnamefont
  {Timoshenko}}\ and\ \bibinfo {author} {\bibfnamefont {S.}~\bibnamefont
  {Woinowsky-Krieger}},\ }\href@noop {} {\emph {\bibinfo {title} {Theory of
  Plates and Shells, 2nd ed}}}\ (\bibinfo  {publisher} {McGraw-Hill, New
  York},\ \bibinfo {year} {1987})\BibitemShut {NoStop}%
\bibitem [{\citenamefont {Bao}\ \emph {et~al.}(2009)\citenamefont {Bao},
  \citenamefont {Miao}, \citenamefont {Chen}, \citenamefont {Zhang},
  \citenamefont {Jang}, \citenamefont {Dames},\ and\ \citenamefont
  {Lau}}]{BaoW2009nn}%
  \BibitemOpen
  \bibfield  {author} {\bibinfo {author} {\bibfnamefont {W.}~\bibnamefont
  {Bao}}, \bibinfo {author} {\bibfnamefont {F.}~\bibnamefont {Miao}}, \bibinfo
  {author} {\bibfnamefont {Z.}~\bibnamefont {Chen}}, \bibinfo {author}
  {\bibfnamefont {H.}~\bibnamefont {Zhang}}, \bibinfo {author} {\bibfnamefont
  {W.}~\bibnamefont {Jang}}, \bibinfo {author} {\bibfnamefont {C.}~\bibnamefont
  {Dames}}, \ and\ \bibinfo {author} {\bibfnamefont {C.~N.}\ \bibnamefont
  {Lau}},\ }\href@noop {} {\bibfield  {journal} {\bibinfo  {journal} {Nature
  Nanotechnology}\ }\textbf {\bibinfo {volume} {4}},\ \bibinfo {pages} {562}
  (\bibinfo {year} {2009})}\BibitemShut {NoStop}%
\bibitem [{\citenamefont {Lu}\ and\ \citenamefont {Huang}(2009)}]{LuQ2009ijam}%
  \BibitemOpen
  \bibfield  {author} {\bibinfo {author} {\bibfnamefont {Q.}~\bibnamefont
  {Lu}}\ and\ \bibinfo {author} {\bibfnamefont {R.}~\bibnamefont {Huang}},\
  }\href@noop {} {\bibfield  {journal} {\bibinfo  {journal} {International
  Journal of Applied Mechanics}\ }\textbf {\bibinfo {volume} {1}},\ \bibinfo
  {pages} {443} (\bibinfo {year} {2009})}\BibitemShut {NoStop}%
\bibitem [{\citenamefont {Patrick}(2010)}]{PatrickWJ2010jctn}%
  \BibitemOpen
  \bibfield  {author} {\bibinfo {author} {\bibfnamefont {W.~J.}\ \bibnamefont
  {Patrick}},\ }\href@noop {} {\bibfield  {journal} {\bibinfo  {journal}
  {Journal of Computational and Theoretical Nanoscience}\ }\textbf {\bibinfo
  {volume} {7}},\ \bibinfo {pages} {2338} (\bibinfo {year} {2010})}\BibitemShut
  {NoStop}%
\bibitem [{\citenamefont {Sakhaee-Pour}(2009)}]{SakhaeePA2009cms}%
  \BibitemOpen
  \bibfield  {author} {\bibinfo {author} {\bibfnamefont {A.}~\bibnamefont
  {Sakhaee-Pour}},\ }\href@noop {} {\bibfield  {journal} {\bibinfo  {journal}
  {Computational Materials Science}\ }\textbf {\bibinfo {volume} {45}},\
  \bibinfo {pages} {266} (\bibinfo {year} {2009})}\BibitemShut {NoStop}%
\bibitem [{\citenamefont {Pradhan}\ and\ \citenamefont
  {Murmu}(2009)}]{PradhanSC2009cms}%
  \BibitemOpen
  \bibfield  {author} {\bibinfo {author} {\bibfnamefont {S.~C.}\ \bibnamefont
  {Pradhan}}\ and\ \bibinfo {author} {\bibfnamefont {T.}~\bibnamefont
  {Murmu}},\ }\href@noop {} {\bibfield  {journal} {\bibinfo  {journal}
  {Computational Materials Science}\ }\textbf {\bibinfo {volume} {47}},\
  \bibinfo {pages} {268} (\bibinfo {year} {2009})}\BibitemShut {NoStop}%
\bibitem [{\citenamefont {Pradhan}(2009)}]{PradhanSC2009plsa}%
  \BibitemOpen
  \bibfield  {author} {\bibinfo {author} {\bibfnamefont {S.~C.}\ \bibnamefont
  {Pradhan}},\ }\href@noop {} {\bibfield  {journal} {\bibinfo  {journal}
  {Physics Letters, Section A: General, Atomic and Solid State Physics}\
  }\textbf {\bibinfo {volume} {373}},\ \bibinfo {pages} {4182} (\bibinfo {year}
  {2009})}\BibitemShut {NoStop}%
\bibitem [{\citenamefont {Frank}\ \emph {et~al.}(2010)\citenamefont {Frank},
  \citenamefont {Tsoukleri}, \citenamefont {Parthenios}, \citenamefont
  {Papagelis}, \citenamefont {Riaz}, \citenamefont {Jalil}, \citenamefont
  {Novoselov},\ and\ \citenamefont {Galiotis}}]{FrankO2010acsnn}%
  \BibitemOpen
  \bibfield  {author} {\bibinfo {author} {\bibfnamefont {O.}~\bibnamefont
  {Frank}}, \bibinfo {author} {\bibfnamefont {G.}~\bibnamefont {Tsoukleri}},
  \bibinfo {author} {\bibfnamefont {J.}~\bibnamefont {Parthenios}}, \bibinfo
  {author} {\bibfnamefont {K.}~\bibnamefont {Papagelis}}, \bibinfo {author}
  {\bibfnamefont {I.}~\bibnamefont {Riaz}}, \bibinfo {author} {\bibfnamefont
  {R.}~\bibnamefont {Jalil}}, \bibinfo {author} {\bibfnamefont {K.~S.}\
  \bibnamefont {Novoselov}}, \ and\ \bibinfo {author} {\bibfnamefont
  {C.}~\bibnamefont {Galiotis}},\ }\href@noop {} {\bibfield  {journal}
  {\bibinfo  {journal} {ACS Nano}\ }\textbf {\bibinfo {volume} {4}},\ \bibinfo
  {pages} {3131} (\bibinfo {year} {2010})}\BibitemShut {NoStop}%
\bibitem [{\citenamefont {Farajpour}\ \emph {et~al.}(2011)\citenamefont
  {Farajpour}, \citenamefont {Mohammadi}, \citenamefont {Shahidi},\ and\
  \citenamefont {Mahzoon}}]{FarajpourA2011pe}%
  \BibitemOpen
  \bibfield  {author} {\bibinfo {author} {\bibfnamefont {A.}~\bibnamefont
  {Farajpour}}, \bibinfo {author} {\bibfnamefont {M.}~\bibnamefont
  {Mohammadi}}, \bibinfo {author} {\bibfnamefont {A.~R.}\ \bibnamefont
  {Shahidi}}, \ and\ \bibinfo {author} {\bibfnamefont {M.}~\bibnamefont
  {Mahzoon}},\ }\href@noop {} {\bibfield  {journal} {\bibinfo  {journal}
  {Physica E: Low-dimensional Systems and Nanostructures}\ }\textbf {\bibinfo
  {volume} {43}},\ \bibinfo {pages} {1820} (\bibinfo {year}
  {2011})}\BibitemShut {NoStop}%
\bibitem [{\citenamefont {Tozzini}\ and\ \citenamefont
  {Pellegrini}(2011)}]{TozziniV2011jpcc}%
  \BibitemOpen
  \bibfield  {author} {\bibinfo {author} {\bibfnamefont {V.}~\bibnamefont
  {Tozzini}}\ and\ \bibinfo {author} {\bibfnamefont {V.}~\bibnamefont
  {Pellegrini}},\ }\href@noop {} {\bibfield  {journal} {\bibinfo  {journal}
  {Journal of Physical Chemistry C}\ }\textbf {\bibinfo {volume} {115}},\
  \bibinfo {pages} {25523} (\bibinfo {year} {2011})}\BibitemShut {NoStop}%
\bibitem [{\citenamefont {Rouhi}\ and\ \citenamefont
  {Ansari}(2012)}]{RouhiS2012pe}%
  \BibitemOpen
  \bibfield  {author} {\bibinfo {author} {\bibfnamefont {S.}~\bibnamefont
  {Rouhi}}\ and\ \bibinfo {author} {\bibfnamefont {R.}~\bibnamefont {Ansari}},\
  }\href@noop {} {\bibfield  {journal} {\bibinfo  {journal} {Physica E:
  Low-dimensional Systems and Nanostructures}\ }\textbf {\bibinfo {volume}
  {44}},\ \bibinfo {pages} {764} (\bibinfo {year} {2012})}\BibitemShut
  {NoStop}%
\bibitem [{\citenamefont {Giannopoulos}(2012)}]{GiannopoulosGI2012cms}%
  \BibitemOpen
  \bibfield  {author} {\bibinfo {author} {\bibfnamefont {G.~I.}\ \bibnamefont
  {Giannopoulos}},\ }\href@noop {} {\bibfield  {journal} {\bibinfo  {journal}
  {Computational Materials Science}\ }\textbf {\bibinfo {volume} {53}},\
  \bibinfo {pages} {388} (\bibinfo {year} {2012})}\BibitemShut {NoStop}%
\bibitem [{\citenamefont {Neek-Amal}\ and\ \citenamefont
  {Peeters}(2012)}]{Neek-AmalM2012apl}%
  \BibitemOpen
  \bibfield  {author} {\bibinfo {author} {\bibfnamefont {M.}~\bibnamefont
  {Neek-Amal}}\ and\ \bibinfo {author} {\bibfnamefont {F.~M.}\ \bibnamefont
  {Peeters}},\ }\href@noop {} {\bibfield  {journal} {\bibinfo  {journal}
  {Applied Physics Letters}\ }\textbf {\bibinfo {volume} {100}},\ \bibinfo
  {pages} {101905} (\bibinfo {year} {2012})}\BibitemShut {NoStop}%
\bibitem [{\citenamefont {Shen}, \citenamefont {Xu},\ and\ \citenamefont
  {Zhang}(2013)}]{ShenH2013apl}%
  \BibitemOpen
  \bibfield  {author} {\bibinfo {author} {\bibfnamefont {H.~.}\ \bibnamefont
  {Shen}}, \bibinfo {author} {\bibfnamefont {Y.~.}\ \bibnamefont {Xu}}, \ and\
  \bibinfo {author} {\bibfnamefont {C.~.}\ \bibnamefont {Zhang}},\ }\href@noop
  {} {\bibfield  {journal} {\bibinfo  {journal} {Applied Physics Letters}\
  }\textbf {\bibinfo {volume} {102}},\ \bibinfo {pages} {131905} (\bibinfo
  {year} {2013})}\BibitemShut {NoStop}%
\bibitem [{\citenamefont {Chang}(2008)}]{ChangT2008prl}%
  \BibitemOpen
  \bibfield  {author} {\bibinfo {author} {\bibfnamefont {T.}~\bibnamefont
  {Chang}},\ }\href@noop {} {\bibfield  {journal} {\bibinfo  {journal}
  {Physical Review Letters}\ }\textbf {\bibinfo {volume} {101}},\ \bibinfo
  {pages} {175501} (\bibinfo {year} {2008})}\BibitemShut {NoStop}%
\bibitem [{\citenamefont {Stoller}\ \emph {et~al.}(2008)\citenamefont
  {Stoller}, \citenamefont {Park}, \citenamefont {Zhu}, \citenamefont {An},\
  and\ \citenamefont {Ruoff}}]{StollerMD2008nl}%
  \BibitemOpen
  \bibfield  {author} {\bibinfo {author} {\bibfnamefont {M.~D.}\ \bibnamefont
  {Stoller}}, \bibinfo {author} {\bibfnamefont {S.}~\bibnamefont {Park}},
  \bibinfo {author} {\bibfnamefont {Y.}~\bibnamefont {Zhu}}, \bibinfo {author}
  {\bibfnamefont {J.}~\bibnamefont {An}}, \ and\ \bibinfo {author}
  {\bibfnamefont {R.~S.}\ \bibnamefont {Ruoff}},\ }\href@noop {} {\bibfield
  {journal} {\bibinfo  {journal} {Nano Letters}\ }\textbf {\bibinfo {volume}
  {8}},\ \bibinfo {pages} {3498} (\bibinfo {year} {2008})}\BibitemShut
  {NoStop}%
\bibitem [{\citenamefont {Wang}\ \emph {et~al.}(2009)\citenamefont {Wang},
  \citenamefont {Shi}, \citenamefont {Huang}, \citenamefont {Ma}, \citenamefont
  {Wang}, \citenamefont {Chen},\ and\ \citenamefont {Chen}}]{WangY2009jpcc}%
  \BibitemOpen
  \bibfield  {author} {\bibinfo {author} {\bibfnamefont {Y.}~\bibnamefont
  {Wang}}, \bibinfo {author} {\bibfnamefont {Z.}~\bibnamefont {Shi}}, \bibinfo
  {author} {\bibfnamefont {Y.}~\bibnamefont {Huang}}, \bibinfo {author}
  {\bibfnamefont {Y.}~\bibnamefont {Ma}}, \bibinfo {author} {\bibfnamefont
  {C.}~\bibnamefont {Wang}}, \bibinfo {author} {\bibfnamefont {M.}~\bibnamefont
  {Chen}}, \ and\ \bibinfo {author} {\bibfnamefont {Y.}~\bibnamefont {Chen}},\
  }\href@noop {} {\bibfield  {journal} {\bibinfo  {journal} {Journal of
  Physical Chemistry C}\ }\textbf {\bibinfo {volume} {113}},\ \bibinfo {pages}
  {131030} (\bibinfo {year} {2009})}\BibitemShut {NoStop}%
\bibitem [{\citenamefont {Liu}\ \emph {et~al.}(2010)\citenamefont {Liu},
  \citenamefont {Yu}, \citenamefont {Neff}, \citenamefont {Zhamu},\ and\
  \citenamefont {Jang}}]{LiuC2010nl}%
  \BibitemOpen
  \bibfield  {author} {\bibinfo {author} {\bibfnamefont {C.}~\bibnamefont
  {Liu}}, \bibinfo {author} {\bibfnamefont {Z.}~\bibnamefont {Yu}}, \bibinfo
  {author} {\bibfnamefont {D.}~\bibnamefont {Neff}}, \bibinfo {author}
  {\bibfnamefont {A.}~\bibnamefont {Zhamu}}, \ and\ \bibinfo {author}
  {\bibfnamefont {B.~Z.}\ \bibnamefont {Jang}},\ }\href@noop {} {\bibfield
  {journal} {\bibinfo  {journal} {Nano Letters}\ }\textbf {\bibinfo {volume}
  {8}},\ \bibinfo {pages} {4863} (\bibinfo {year} {2010})}\BibitemShut
  {NoStop}%
\bibitem [{\citenamefont {Weng}\ \emph {et~al.}(2011)\citenamefont {Weng},
  \citenamefont {Su}, \citenamefont {Wang}, \citenamefont {Li}, \citenamefont
  {Du},\ and\ \citenamefont {Cheng}}]{WangZ2011adm}%
  \BibitemOpen
  \bibfield  {author} {\bibinfo {author} {\bibfnamefont {Z.}~\bibnamefont
  {Weng}}, \bibinfo {author} {\bibfnamefont {Y.}~\bibnamefont {Su}}, \bibinfo
  {author} {\bibfnamefont {D.-W.}\ \bibnamefont {Wang}}, \bibinfo {author}
  {\bibfnamefont {F.}~\bibnamefont {Li}}, \bibinfo {author} {\bibfnamefont
  {J.}~\bibnamefont {Du}}, \ and\ \bibinfo {author} {\bibfnamefont {H.-M.}\
  \bibnamefont {Cheng}},\ }\href@noop {} {\bibfield  {journal} {\bibinfo
  {journal} {Advanced Materials}\ }\textbf {\bibinfo {volume} {1}},\ \bibinfo
  {pages} {917} (\bibinfo {year} {2011})}\BibitemShut {NoStop}%
\bibitem [{\citenamefont {Brownson}, \citenamefont {Kampouris},\ and\
  \citenamefont {Banks}(2011)}]{BrownsonDAC2011jps}%
  \BibitemOpen
  \bibfield  {author} {\bibinfo {author} {\bibfnamefont {D.~A.}\ \bibnamefont
  {Brownson}}, \bibinfo {author} {\bibfnamefont {D.~K.}\ \bibnamefont
  {Kampouris}}, \ and\ \bibinfo {author} {\bibfnamefont {C.~E.}\ \bibnamefont
  {Banks}},\ }\href@noop {} {\bibfield  {journal} {\bibinfo  {journal} {Journal
  of Power Source}\ }\textbf {\bibinfo {volume} {196}},\ \bibinfo {pages}
  {4873} (\bibinfo {year} {2011})}\BibitemShut {NoStop}%
\bibitem [{\citenamefont {Brenner}\ \emph {et~al.}(2002)\citenamefont
  {Brenner}, \citenamefont {Shenderova}, \citenamefont {Harrison},
  \citenamefont {Stuart}, \citenamefont {Ni},\ and\ \citenamefont
  {Sinnott}}]{brennerJPCM2002}%
  \BibitemOpen
  \bibfield  {author} {\bibinfo {author} {\bibfnamefont {D.~W.}\ \bibnamefont
  {Brenner}}, \bibinfo {author} {\bibfnamefont {O.~A.}\ \bibnamefont
  {Shenderova}}, \bibinfo {author} {\bibfnamefont {J.~A.}\ \bibnamefont
  {Harrison}}, \bibinfo {author} {\bibfnamefont {S.~J.}\ \bibnamefont
  {Stuart}}, \bibinfo {author} {\bibfnamefont {B.}~\bibnamefont {Ni}}, \ and\
  \bibinfo {author} {\bibfnamefont {S.~B.}\ \bibnamefont {Sinnott}},\
  }\href@noop {} {\bibfield  {journal} {\bibinfo  {journal} {Journal of
  Physics: Condensed Matter}\ }\textbf {\bibinfo {volume} {14}},\ \bibinfo
  {pages} {783} (\bibinfo {year} {2002})}\BibitemShut {NoStop}%
\bibitem [{\citenamefont {Nose}(1984)}]{Nose}%
  \BibitemOpen
  \bibfield  {author} {\bibinfo {author} {\bibfnamefont {S.}~\bibnamefont
  {Nose}},\ }\href@noop {} {\bibfield  {journal} {\bibinfo  {journal} {Journal
  of Chemical Physics}\ }\textbf {\bibinfo {volume} {81}},\ \bibinfo {pages}
  {511} (\bibinfo {year} {1984})}\BibitemShut {NoStop}%
\bibitem [{\citenamefont {Hoover}(1985)}]{Hoover}%
  \BibitemOpen
  \bibfield  {author} {\bibinfo {author} {\bibfnamefont {W.~G.}\ \bibnamefont
  {Hoover}},\ }\href@noop {} {\bibfield  {journal} {\bibinfo  {journal}
  {Physical Review A}\ }\textbf {\bibinfo {volume} {31}},\ \bibinfo {pages}
  {1695} (\bibinfo {year} {1985})}\BibitemShut {NoStop}%
\bibitem [{\citenamefont {Plimpton}(1995)}]{PlimptonSJ}%
  \BibitemOpen
  \bibfield  {author} {\bibinfo {author} {\bibfnamefont {S.~J.}\ \bibnamefont
  {Plimpton}},\ }\href@noop {} {\bibfield  {journal} {\bibinfo  {journal}
  {Journal of Computational Physics}\ }\textbf {\bibinfo {volume} {117}},\
  \bibinfo {pages} {1} (\bibinfo {year} {1995})}\BibitemShut {NoStop}%
\bibitem [{\citenamefont {Stukowski}(2010)}]{ovito}%
  \BibitemOpen
  \bibfield  {author} {\bibinfo {author} {\bibfnamefont {A.}~\bibnamefont
  {Stukowski}},\ }\href@noop {} {\bibfield  {journal} {\bibinfo  {journal}
  {Modelling and Simulation in Materials Science and Engineering}\ }\textbf
  {\bibinfo {volume} {18}},\ \bibinfo {pages} {015012} (\bibinfo {year}
  {2010})}\BibitemShut {NoStop}%
\bibitem [{\citenamefont {Green}\ \emph {et~al.}(2014)\citenamefont {Green},
  \citenamefont {Emery}, \citenamefont {Hishikawa}, \citenamefont {Warta},\
  and\ \citenamefont {Dunlop}}]{GreenMA2014ppra}%
  \BibitemOpen
  \bibfield  {author} {\bibinfo {author} {\bibfnamefont {M.~A.}\ \bibnamefont
  {Green}}, \bibinfo {author} {\bibfnamefont {K.}~\bibnamefont {Emery}},
  \bibinfo {author} {\bibfnamefont {Y.}~\bibnamefont {Hishikawa}}, \bibinfo
  {author} {\bibfnamefont {W.}~\bibnamefont {Warta}}, \ and\ \bibinfo {author}
  {\bibfnamefont {E.~D.}\ \bibnamefont {Dunlop}},\ }\href@noop {} {\bibfield
  {journal} {\bibinfo  {journal} {Prog. Photovolt: Res. Appl.}\ }\textbf
  {\bibinfo {volume} {22}},\ \bibinfo {pages} {701} (\bibinfo {year}
  {2014})}\BibitemShut {NoStop}%
\bibitem [{\citenamefont {Chen}\ \emph {et~al.}(2009)\citenamefont {Chen},
  \citenamefont {Rosenblatt}, \citenamefont {Bolotin}, \citenamefont {Kalb},
  \citenamefont {Kim}, \citenamefont {Kymissis}, \citenamefont {Stormer},
  \citenamefont {Heinz},\ and\ \citenamefont {Hone}}]{ChenC2009nn}%
  \BibitemOpen
  \bibfield  {author} {\bibinfo {author} {\bibfnamefont {C.}~\bibnamefont
  {Chen}}, \bibinfo {author} {\bibfnamefont {S.}~\bibnamefont {Rosenblatt}},
  \bibinfo {author} {\bibfnamefont {K.~I.}\ \bibnamefont {Bolotin}}, \bibinfo
  {author} {\bibfnamefont {W.}~\bibnamefont {Kalb}}, \bibinfo {author}
  {\bibfnamefont {P.}~\bibnamefont {Kim}}, \bibinfo {author} {\bibfnamefont
  {I.}~\bibnamefont {Kymissis}}, \bibinfo {author} {\bibfnamefont {H.~L.}\
  \bibnamefont {Stormer}}, \bibinfo {author} {\bibfnamefont {T.~F.}\
  \bibnamefont {Heinz}}, \ and\ \bibinfo {author} {\bibfnamefont
  {J.}~\bibnamefont {Hone}},\ }\href@noop {} {\bibfield  {journal} {\bibinfo
  {journal} {Nature Nanotechnology}\ }\textbf {\bibinfo {volume} {4}},\
  \bibinfo {pages} {861} (\bibinfo {year} {2009})}\BibitemShut {NoStop}%
\bibitem [{\citenamefont {van~der Zande}\ \emph {et~al.}(2010)\citenamefont
  {van~der Zande}, \citenamefont {Barton}, \citenamefont {Alden}, \citenamefont
  {Ruiz-Vargas}, \citenamefont {Whitney}, \citenamefont {Pham}, \citenamefont
  {Park}, \citenamefont {Parpia}, \citenamefont {Craighead},\ and\
  \citenamefont {McEuen}}]{ZandeAMVD}%
  \BibitemOpen
  \bibfield  {author} {\bibinfo {author} {\bibfnamefont {A.~M.}\ \bibnamefont
  {van~der Zande}}, \bibinfo {author} {\bibfnamefont {R.~A.}\ \bibnamefont
  {Barton}}, \bibinfo {author} {\bibfnamefont {J.~S.}\ \bibnamefont {Alden}},
  \bibinfo {author} {\bibfnamefont {C.~S.}\ \bibnamefont {Ruiz-Vargas}},
  \bibinfo {author} {\bibfnamefont {W.~S.}\ \bibnamefont {Whitney}}, \bibinfo
  {author} {\bibfnamefont {P.~H.~Q.}\ \bibnamefont {Pham}}, \bibinfo {author}
  {\bibfnamefont {J.}~\bibnamefont {Park}}, \bibinfo {author} {\bibfnamefont
  {J.~M.}\ \bibnamefont {Parpia}}, \bibinfo {author} {\bibfnamefont {H.~G.}\
  \bibnamefont {Craighead}}, \ and\ \bibinfo {author} {\bibfnamefont {P.~L.}\
  \bibnamefont {McEuen}},\ }\href@noop {} {\bibfield  {journal} {\bibinfo
  {journal} {Nano Letters}\ }\textbf {\bibinfo {volume} {10}},\ \bibinfo
  {pages} {4869} (\bibinfo {year} {2010})}\BibitemShut {NoStop}%
\bibitem [{\citenamefont {Garcia-Sanchez}\ \emph {et~al.}(2008)\citenamefont
  {Garcia-Sanchez}, \citenamefont {van~der Zande}, \citenamefont {Paulo},
  \citenamefont {Lassagne}, \citenamefont {McEuen},\ and\ \citenamefont
  {Bachtold}}]{SanchezDG}%
  \BibitemOpen
  \bibfield  {author} {\bibinfo {author} {\bibfnamefont {D.}~\bibnamefont
  {Garcia-Sanchez}}, \bibinfo {author} {\bibfnamefont {A.~M.}\ \bibnamefont
  {van~der Zande}}, \bibinfo {author} {\bibfnamefont {A.~S.}\ \bibnamefont
  {Paulo}}, \bibinfo {author} {\bibfnamefont {B.}~\bibnamefont {Lassagne}},
  \bibinfo {author} {\bibfnamefont {P.~L.}\ \bibnamefont {McEuen}}, \ and\
  \bibinfo {author} {\bibfnamefont {A.}~\bibnamefont {Bachtold}},\ }\href@noop
  {} {\bibfield  {journal} {\bibinfo  {journal} {Nano Letters}\ }\textbf
  {\bibinfo {volume} {8}},\ \bibinfo {pages} {1399} (\bibinfo {year}
  {2008})}\BibitemShut {NoStop}%
\end{thebibliography}
\end{document}